\documentclass[fleqn,twoside]{article}
\usepackage{espcrc2}

\usepackage{graphicx}
\usepackage[figuresright]{rotating}

\newcommand{\AmS}{{\protect\the\textfont2
  A\kern-.1667em\lower.5ex\hbox{M}\kern-.125emS}}

\title{MSSM Higgs searches with tau lepton final states in CMS}

\author{S. Gennai\address[MCSD]{Scuola Normale Superiore and Istituto Nazionale di Fisica 
Nucleare Pisa, 
    Italy.}
 }
  
\begin{document}
    				   
\begin{abstract}

The present understanding of large mass MSSM Higgs sector is reviewed.
The most profitable channels: A/H~$\rightarrow \tau \tau$ and $H^{\pm} 
\rightarrow \tau \nu$ are considered;  
a glimpse of the trigger
chain studied for events with tau lepton final states is presented.
The MSSM Higgs discovery reach of the general-purpouse experiment CMS is 
summarised.     
\vspace{1pc}
\end{abstract}

\maketitle

\section{Introduction}
The Standard Model~\cite{intro:sm} of strong and electroweak interactions 
is in excellent agreement with the experimental
measurements. However, the core of the theory, the electroweak symmetry 
breaking manifesting itself 
in the heavy vector bosonsW and Z and the massless photon, is the least known 
sector of the model. The Higgs mechanism~\cite{intro:higgsmech} provides a 
mathematical explanation to this phenomenon, 
and one of the main tasks of the LHC collider will be
the quest of the Higgs particle experimental evidence, or any observable of
 some other symmetry breaking
mechanism.
Without the Higgs boson the Standard Model is neither consistent nor complete, since 
the masses of the gauge bosons
and fermions are generated through the interaction with the Higgs field.\\
The only unknown parameter in the SM Higgs sector is the mass of the Higgs boson. 
This is not predicted by
the theory, but indirect constraints for the possible mass range can be deduced 
from theoretical 
arguments~\cite{theo:higgs}.
Furthermore, the electroweak precision measurements, where the Higgs mass enters 
in the radiative corrections,
can be used to predict the most likely value of the Higgs mass consistent with 
all the experimental data 
used in the fit. Such fits favour a rather light Higgs boson, $m_{H} <$ 196 GeV
 with 95\% confidence 
level~\cite{lep:mh1}. Direct searches at LEP exclude the SM Higgs 
boson below $m_{H}$ = 
114.1 GeV at 95\% confidence level~\cite{lep:mh2}.
The SM Higgs would be the first fundamental scalar particle.
However, when the bare mass of this scalar 
particle is computed in the 
perturbation theory, it turns out that the mass diverges quadratically. 
Technically, this problem could 
be solved by renormalization, resulting in a counter-term balancing the quadratic
 divergence in each order 
of the perturbative calculations, but such fine-tuning cannot be considered
 natural or elegant. 
This unpleasant feature of the SM is one of the main motivations to search
for a theory without such a drawback as supersimmetric theories~\cite{mssm:theo}. 
In this report, the Higgs sector of one of these supersymmetric theories is reviewed.
The production mechanism at the LHC are briefly discussed as well as the possible 
decay modes. The discovery potential of the general-purpose detector CMS~\cite{intro:cms} is 
summarised without going into details of the detector performance. 
\subsection{The MSSM Higgs sector}
In supersymmetric theories, for each SM particle a 
supersymmetric partner is introduced. These sparticles have
the same quantum numbers as the particles but their spin differs by one half. 
The introduction of the supersymmetric
partners cancels the quadratic divergence in the Higgs boson mass, thus solving 
the fine-tuning problem,
provided the masses of the supersymmetric partners are not beyond 1 TeV scale.
In this report, we concentrate on the Minimal Supersymmetric 
extesion~\cite{mssm:theo1} of the 
Standard Model, which is minimal
in the sense that a minimum number, i.e. two, of Higgs doublets is introduced. 
This results in five observable Higgs
particles in the MSSM: two neutral CP-even scalars, a light h and a heavy H, 
a CP-odd A, and charged $H^+$ and
$H^-$. 
In the MSSM, at tree level, the Higgs sector is defined by two parameters 
which can be chosen to be the mass
of the CP-odd A, $m_{A}$ and $\tan \beta$, the ratio of the vacuum 
expectation values of 
the two Higgs doublets. 
There are other parameters which affect the Higgs sector through radiative corrections, 
such as the top quark mass, the mass
scales of the SUSY particles and the mixing between the left and right handed 
components of the stop squark.
The two parameters, $m_{A}$ and $\tan \beta$, 
define the masses of other Higgs particles ~\cite{mssm:theo2}. 
The light h reaches its maximal mass 
already at moderate $m_{A}$ values. Above $m_A \sim$ 200 GeV the 
heavy Higgses (H, A and $H^{+}$) are almost degenerate in mass. 
The LEP experiments have excluded a light h below 91 GeV, an A below 91.9 GeV 
and a charged Higgs below 78.6 GeV~\cite{lep:mssm}. 
For maximal stop mixing the range 0.5 $< \tan \beta <$ 2.4 has been excluded, 
and for minimal mixing
0.7 $< \tan \beta <$ 10.5~\cite{lep:mssm}.\\ 
In this report we concentrate on large mass A/H and $H^+$, decaying in tau lepton 
which are the most promising channels investigated till now.
\section{Higgs bosons production and decay at LHC}
\begin{figure}[ht]
\vspace{9pt}
\includegraphics[width=13pc]{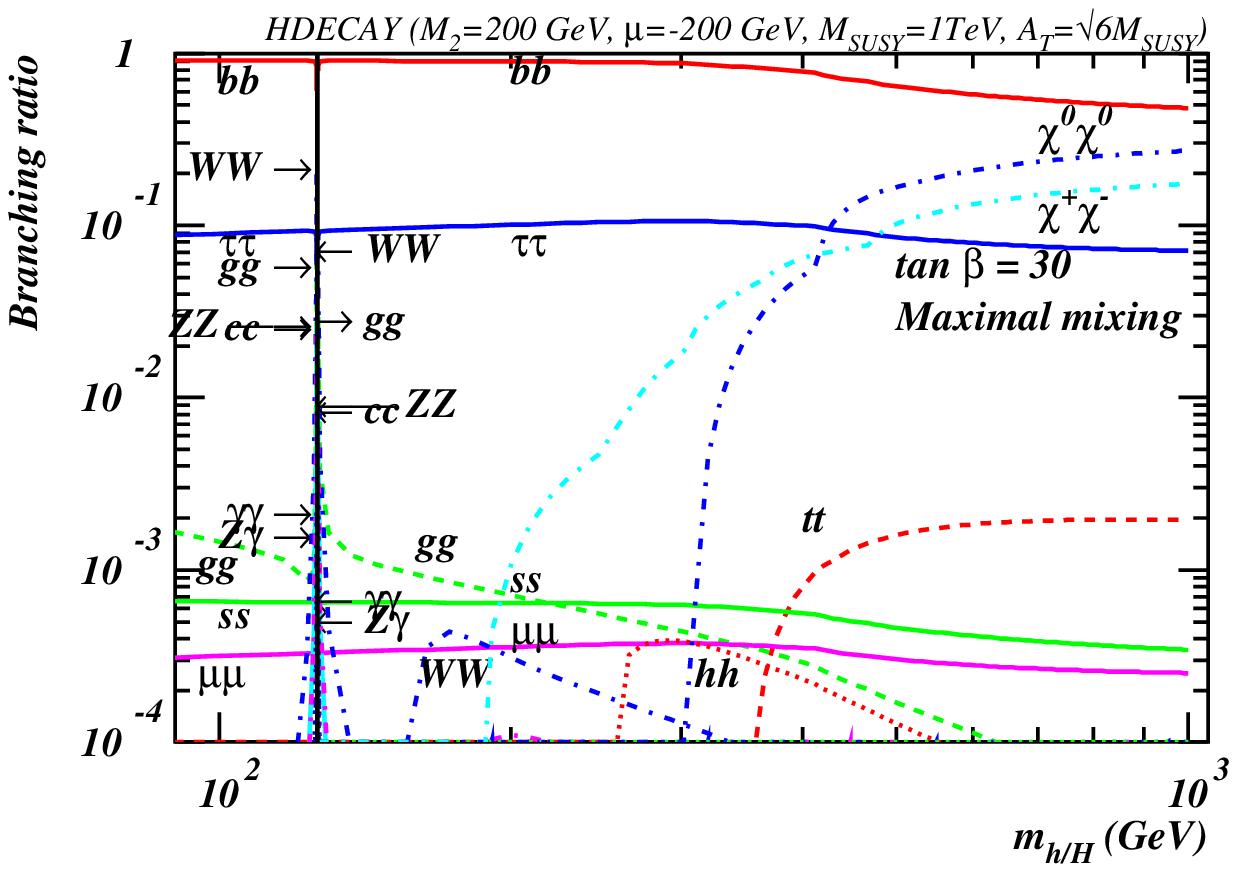}
\includegraphics[width=13pc]{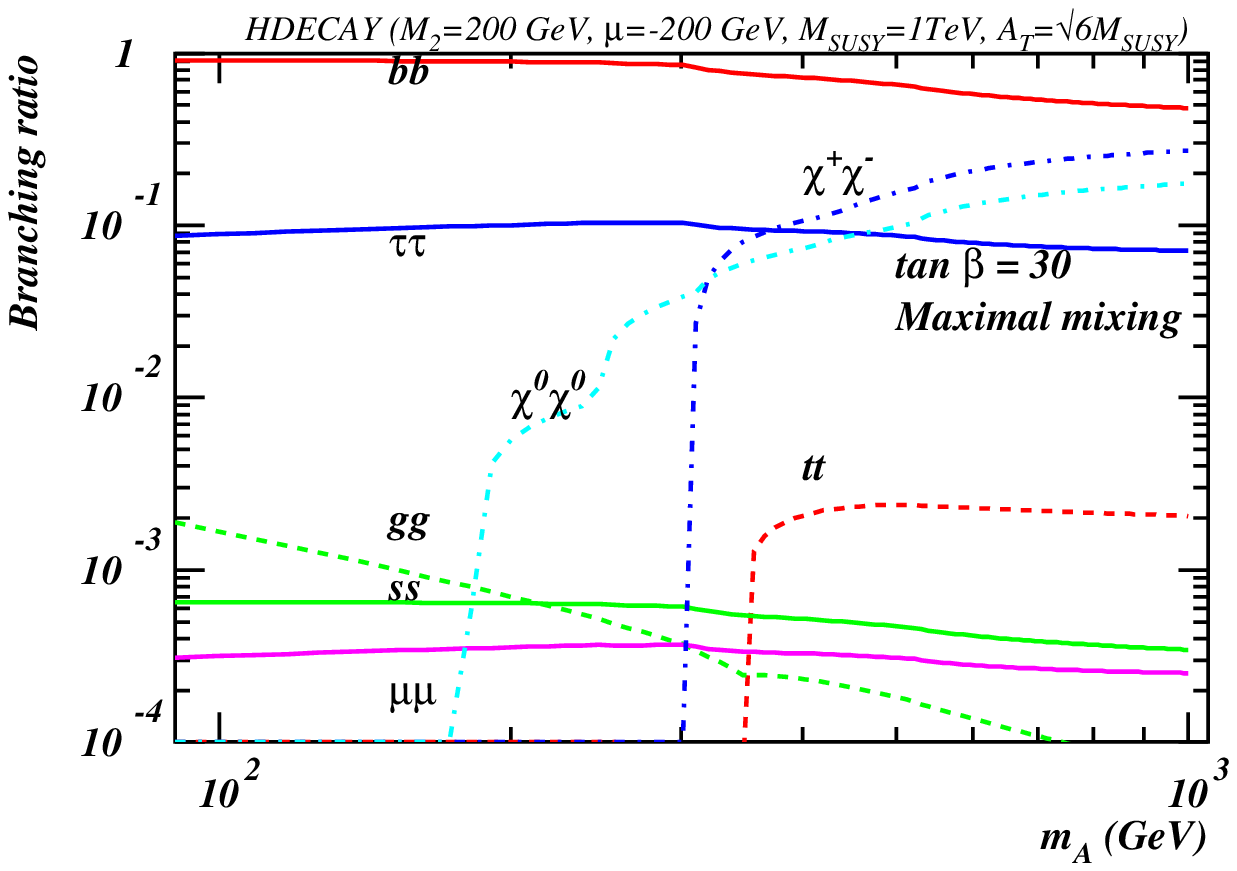}
\caption{The upper plots is the Higgs 
branching ratios of the light h and heavy H 
as a function of their masses. The lower one is for the pseudo-scalar Higgs A
as a function of $m_{A}$.}
\label{fig:brh}
\end{figure}

\begin{figure}[ht]
\vspace{9pt}
\includegraphics[width=13pc]{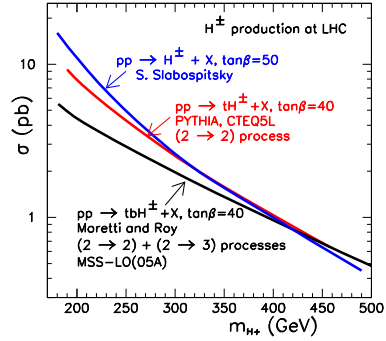}
\includegraphics[width=13pc]{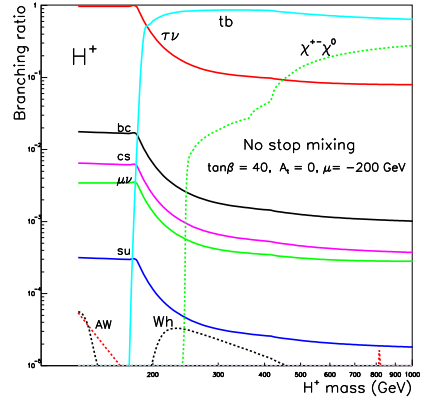}
\caption{The upper plot is the charged Higgs production cross section 
as a function of $m_{H^{+}}$. The lower one is the charged Higgs 
branching ratios as a function of $m_{H^+}$.}
\label{fig:sigmahp}
\end{figure}

In this section, Higgs production and decay at the LHC are reviewed. 
The cross-sections have been computed with the Higgs production programs HIGLU, 
VV2H, V2HV and HQQ~\cite{theo:cross} based on the calculations in~\cite{theo:spira}
 and the branching ratios have been computed with HDECAY~\cite{theo:decay}.\\
\begin{figure}[ht]
\vspace{9pt}
\includegraphics[width=14pc]{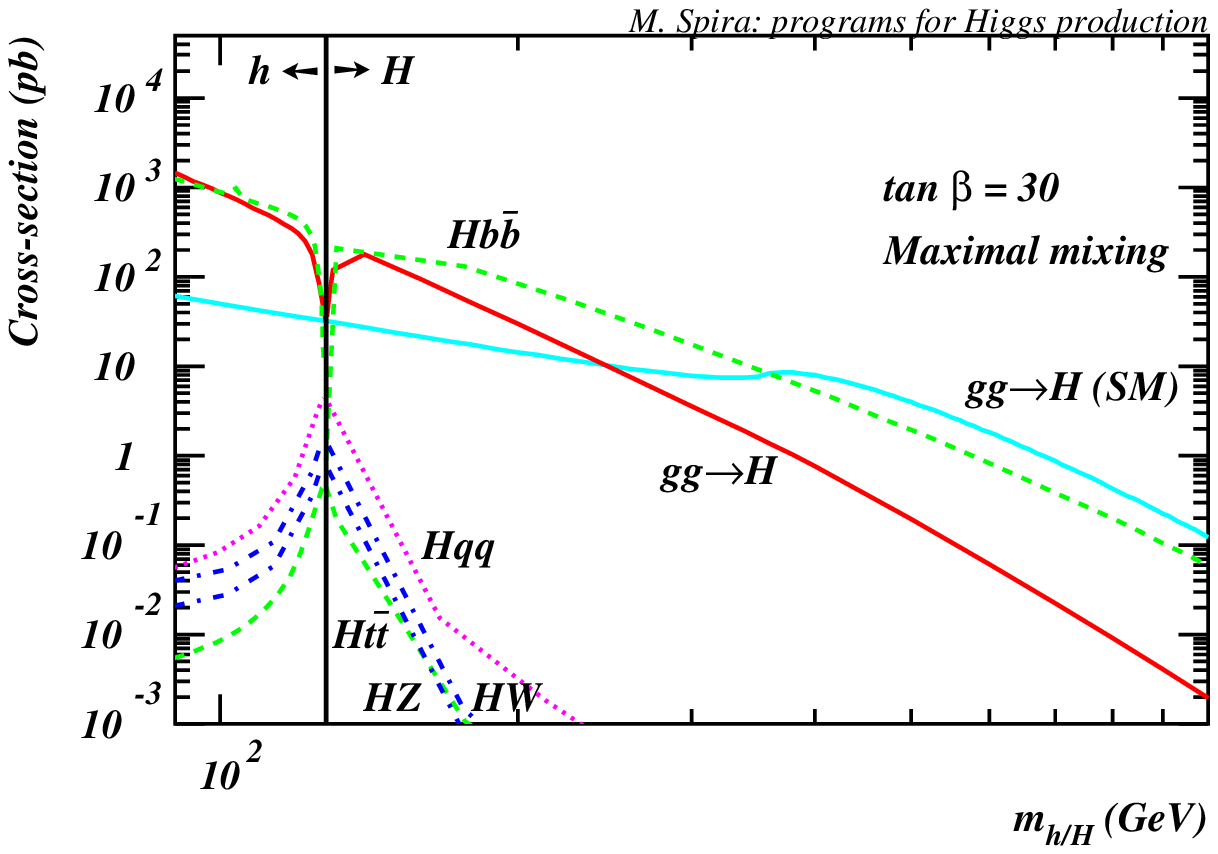}
\includegraphics[width=14pc]{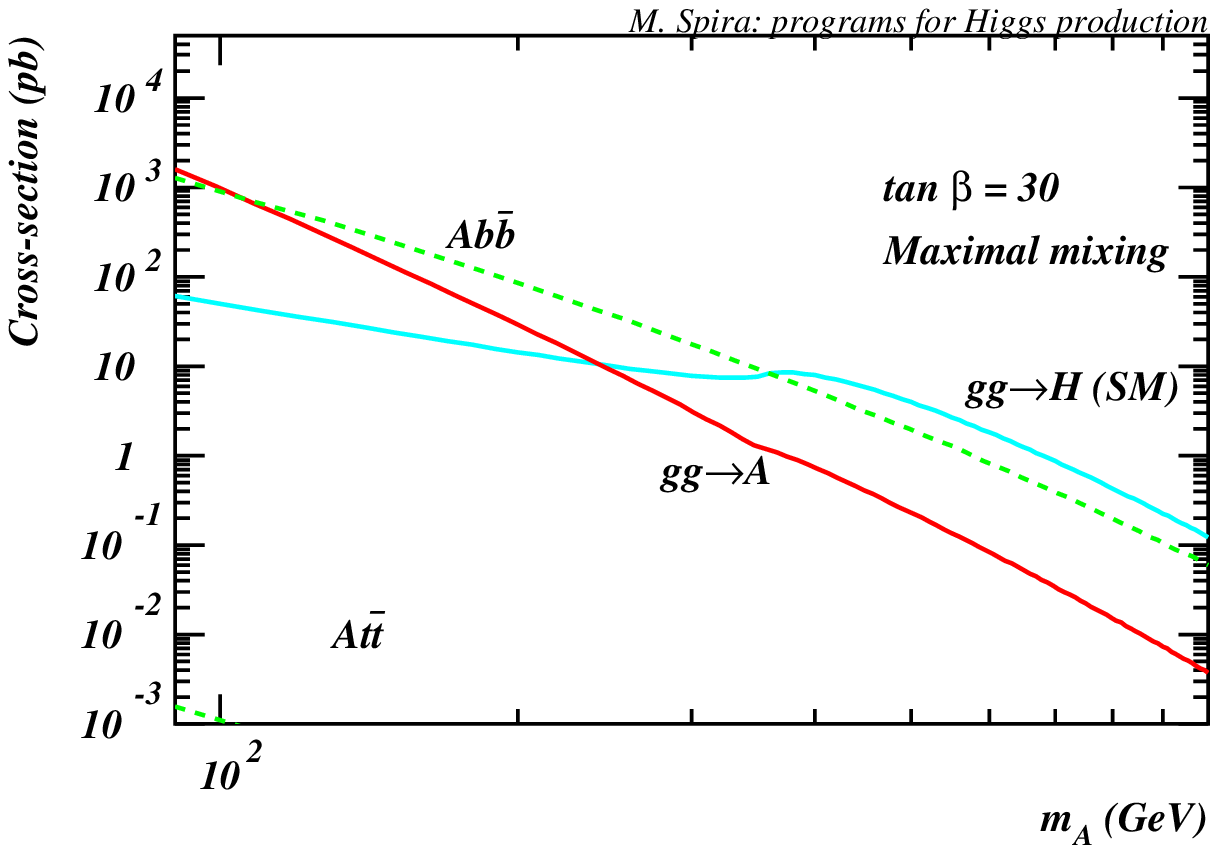}
\caption{The upper plots is the Higgs production cross-section of the light h and heavy H 
as a function of their masses. 
The lower one is Pseudo-scalar Higgs production cross-section  
as a function of $m_{A}$.}
\label{fig:massh}
\end{figure}
In the MSSM, the SM Higgs couplings are modified as a function of the $m_A$ and $\tan \beta$ 
at tree level. The resulting cross-sections are shown for $\tan \beta$ = 30 in 
figure~\ref{fig:massh} for h,H and A. 
For comparison, the dominant SM Higgs production process through the gluon-gluon fusion, 
is shown in the plots.
It can be concluded that the total rate is enhanced, respect to SM value, 
at high $\tan \beta$ and $m_A <$ 400 GeV. The vector
boson fusion is suppressed, for h and H especially at high $\tan \beta$, 
and altogether for A which does not couple to
vector bosons at tree level. 
Higgs production in association with a $b\overline{b}$ pair is strongly enhanced and it becomes the
dominant production mechanism at high $\tan \beta$.
The Higgs decay pattern can be extremely complicated as shown in figure~\ref{fig:brh} 
for h and H. For the light h, the
branching ratios reach their SM value when $m_h$ reaches its maximum value. 
It is worth noting that even if this area
is just a narrow line when plotted as a function of $m_h$, 
it covers most of the $m_A - \tan \beta$ parameter plane. 
The decay into a $b \overline{b}$ pair is dominant for the light h with 
$m_h < m_{h}^{max}$ and for the heavy H at high $\tan \beta$. 
The same figure shows the branching ratios for A. 
The $b \overline{b}$ decay is dominant and the $\tau \tau$ 
decay much more significant than
in the SM where the opening of the vector boson channels suppresses the fermion decays. 
As illustrated in the
plot, the decays to SUSY particle may be important if their masses are light enough.\\
The charged Higgs can be produced in the top quark decay if it mass is lighter than the 
top quark mass. If it is heavier, it is produced in other processes alone, or in association 
with a top quark or a tb quark pair, see figure~\ref{fig:sigmahp} where also the branching
ratios of $H^{+}$ decays are shown.
The decay to a tb quark pair is dominant where 
kinematically possible.
Below top quark mass, 
the $\tau \nu$ decay is dominant. The SUSY parameters chosen for the plot allow 
the decay to a chargino neutralino pair in the high mass range.
From the values of the branching ratios it can be seen that the  channels with 
$b$ quarks in the final states are favoured. However from more detailed studies it turned out that
the channel with $\tau$ final states have bigger discovery potential, thanks to the cleaner signal
and lower background.
\section{Higgs searches}
\begin{figure}[ht]
\vspace{9pt}
\includegraphics[width=12pc]{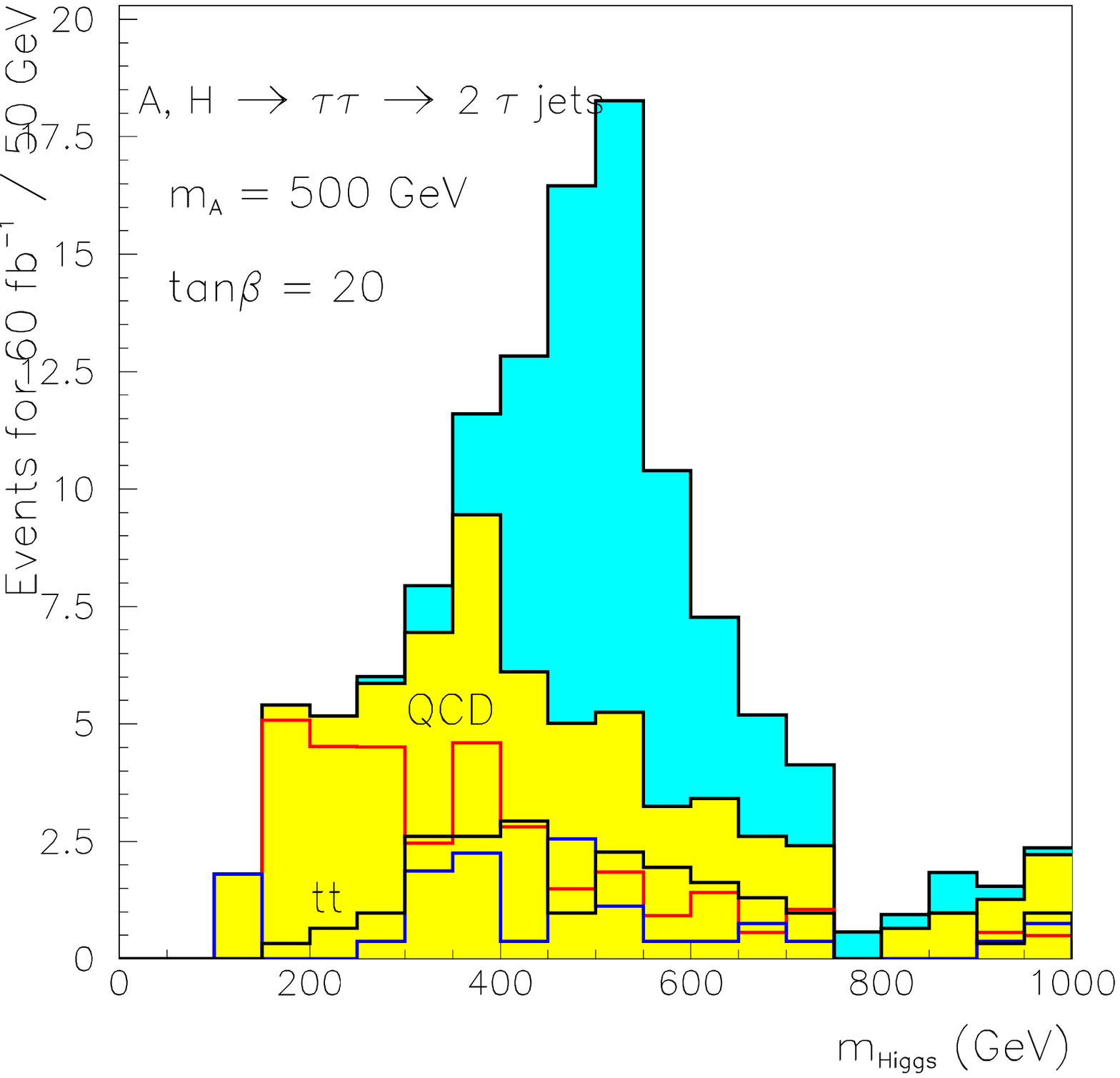}
\includegraphics[width=11pc]{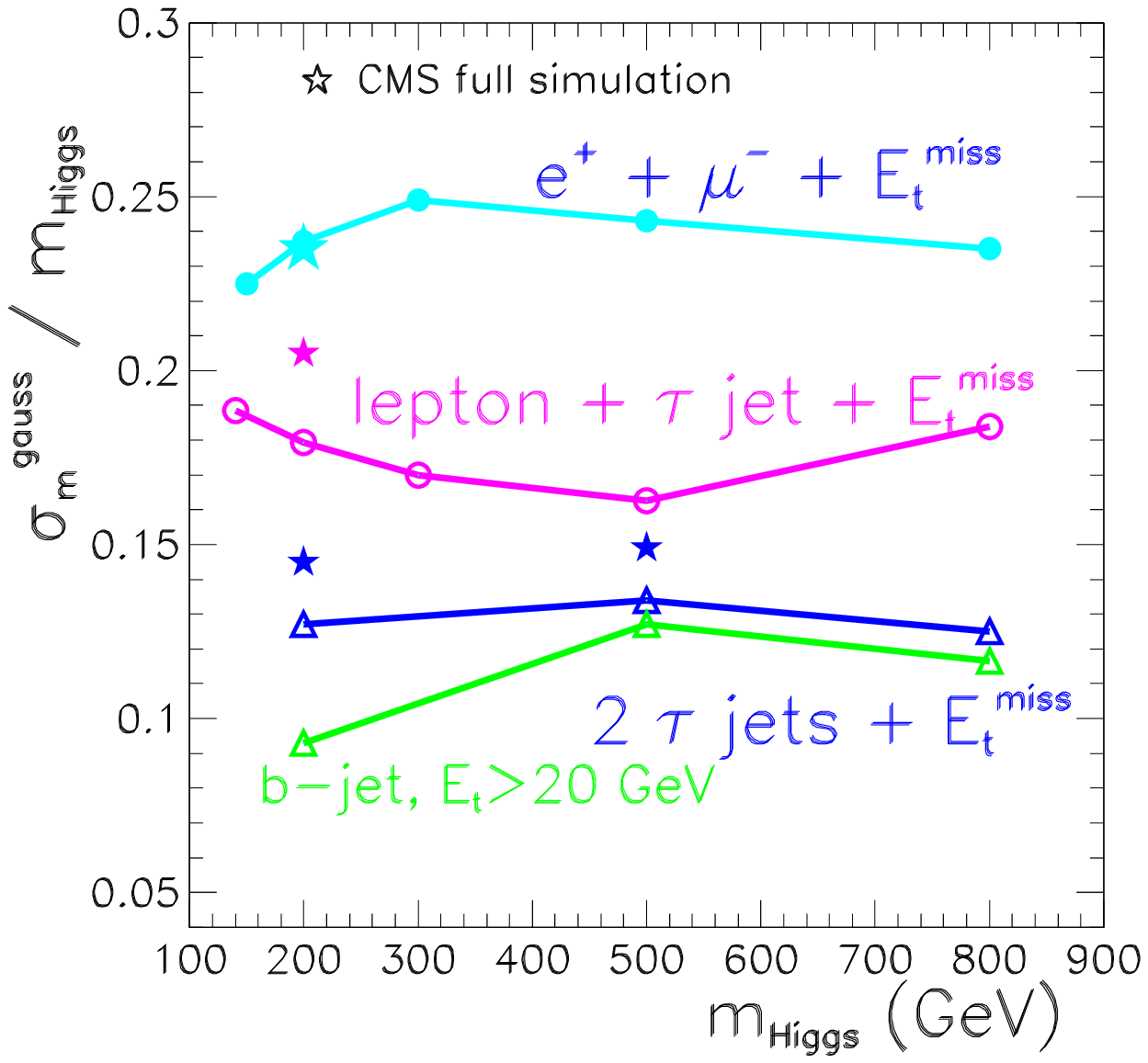}
\caption{The upper plot is the reconstructed mass in the 
$H/A \rightarrow \tau \tau \rightarrow 2 \tau_{jet}$ channel,
for a Higgs mass of 500 GeV (the main background is superimposed).
The lower one is the mass resolution for the several studied channels.}
\label{fig:hmass}
\end{figure}

One of the aim of LHC is to cover the entire $m_A - \tan \beta$ plane in order to discover or 
exclude the esistence of the MSSM Higgs sector. 
Among the several MSSM decay modes two channels containing tau lepton final states 
are of great importance $H/A \rightarrow \tau \tau$ and
$H^{+} \rightarrow \tau \nu$, which have been studied in detailed. 
These channels will be shortly discussed in the following. 
There are several other channels to study different regions of the parameter space, 
such as heavy H or A decaying in light h, 
muonic decay of H or A and tb pair decay of the charged Higgs.
Furthermore, if the SUSY mass scale allows the decay into SUSY particles the $H/A \rightarrow
\chi^{0}_{2} \chi^{0}_{2}$
could be visible. \\
However $H/A \rightarrow \tau \tau$  decay~\cite{mssm:htautau} is the most promising channel 
in the search for the neutral heavy Higgses, and it is
particularly significant at high $\tan \beta$. 
Experimentally, the $\tau$ decays are very interesting, they require an interplay
of different detector elements measuring missing $E_T$ due to the escaping neutrinos, 
leptons from the leptonic $\tau$
decay, jets from the hadronic $\tau$ decays. Furthermore some b-tagging could be required in order
to suppress the background when b-quarks are produced in association with the Higgs.
Indeed the best significance for this channel is obtained with
the Higgs production in association with a $b \overline{b}$ pair. 
Three final states have been considered: $\tau \tau \rightarrow \l \l + \nu's$, 
$\tau \tau \rightarrow \tau_{jet} \l + \nu's$ and 
$\tau \tau \rightarrow \tau_{jet} \tau_{jet} + \nu's$. 
The event with two $\tau_{jet}$ can be triggered with a specific $\tau$ 
trigger~\cite{trigger:tau}, while the other can be triggered  with combiuned electron
and muon trigger as described in the next section. 
The main backgrounds for this channel are represented by:
\begin{itemize}
\item Z,$\gamma^* \rightarrow \tau \tau$
\item $t\overline{t}$
\item W,Z+jets
\item QCD jets
\end{itemize}
The QCD background, the one with the biggest cross section, 
is effectively suppressed by the isolation criteria and the missing
$E_T$ cut.
Isolation criteria applied at a high level trigger on the total 
hadronic $\tau$ decays can reduce QCD contamination by a 
factor of 1000 when applied to both jets.   
Furthermore, the leptonic $\tau$ decays can be identified with $\tau$ tagging 
using an impact parameter cut. 
Assuming the neutrinos collinear to the $\tau_{jet}$ directions is possible to reconstruct
the invariant mass of the final state. 
The reconstructed mass in the $H/A \rightarrow \tau \tau \rightarrow 2 \tau_{jet}$ channel is 
shown in figure~\ref{fig:hmass}~\cite{mssm:hmass}, for a Higgs mass of 500 GeV (the main 
background is superimposed). 
As shown in the plot the signal can be clearly extratcted fro background with a fit to 
the reconstructed mass.
\begin{figure}[ht]
\vspace{9pt}
\includegraphics[width=12pc]{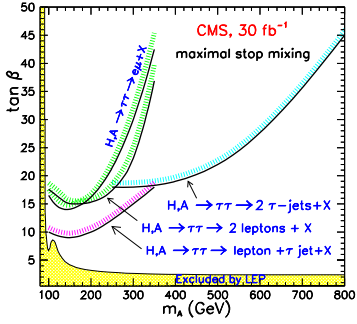}
\caption{Expected 5 sigma discovery region in the $m_A - \tan \beta$ plane for the heavy 
Higgs H and A with the $\tau \tau$ 
final state for 30 $fb^{-1}$.}
\label{fig:5sigma}
\end{figure}
\begin{figure}[ht]
\vspace{9pt}
\includegraphics[height=10pc]{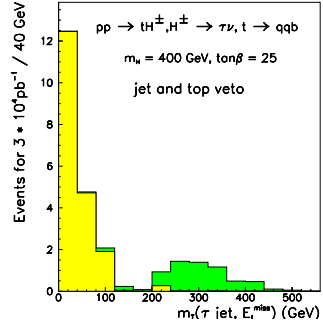}
\caption{Transverse mass reconstruction from $\tau_{jet}$ and missing $E_{T}$
for $pp \rightarrow tH^{\pm}, H^{\pm} \rightarrow \tau \nu$ for a $m_{H^{\pm}}$=400 GeV
and $\tan \beta =30$. The total background is superimposed with a veto for a second top and
a central jet in the event.}
\label{fig:hpmass}
\end{figure}
\begin{figure}[ht]
\vspace{8pt}
\includegraphics[width=12pc]{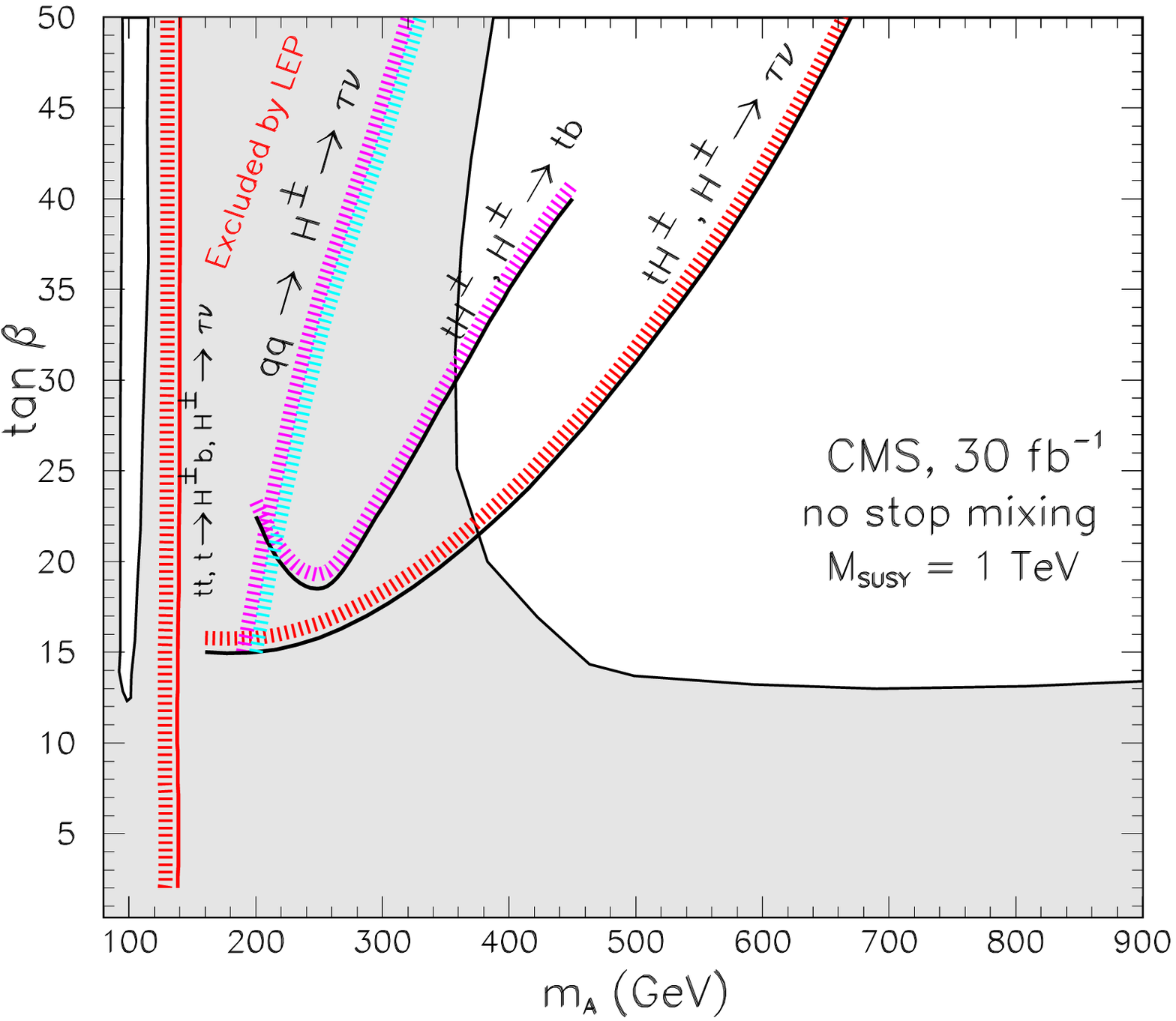}
\caption{Expected 5 sigma discovery region in the $m_A - \tan \beta$ plane for the 
$H^{+} \rightarrow \tau \tau$ and  $H^{+} \rightarrow tb$ for 30 $fb^{-1}$ for 
maximal stop mixing.}
\label{fig:hp5sigma}
\end{figure}
The same figure~\cite{mssm:hmass} shows the mass resolution for the different 
considered channels. The best resolution is obtained with the two $\tau_{jet}$ final state 
when using also b-tagging.
The parameter space coverage for the heavy Higgs H and A is shown in 
figure~\ref{fig:5sigma}~\cite{mssm:hmass}. 
The $\tau$ decay channels cover the high $tan \beta$ part of the
parameter space. 
The discovery range for these channels extend down to $\tan \beta \sim$15 at $m_A$ = 300 GeV and
down to   $\tan \beta \sim$20 at $m_A$ = 500 GeV
The middle $\tan \beta$ values are not covered by these channels, other possibilities
involving SUSY sparticle decays are under study~\cite{mssm:ritva}.
The charged Higgs can be observed in the $t \overline{t}$ events if $m_{H^{\pm}} < m_{top}$ 
or in the decay into $\tau \nu$ when $m_{H^{\pm}} > m_{top}$
and $H^{\pm}$ is produced in association with a top quark.
In this report we will focus on the $m_{H^{\pm}} > m_{top}$ considering the $\tau \nu$ 
final state, while only total hadronic top quark decay
will be considered.
Figure~\ref{fig:hpmass} shows the reconstructed transverse mass for $m_{H^{\pm}}$
= 400 GeV  with expected background for 30 $fb^{-1}$~\cite{mssm:kinn}.\\
To obtain a clear separation between the
signal and the background the $\tau$ polarisation properties has been exploited. 
The polarisation of a $\tau$ depend on the spin of the mother particle.
The hadronic decay products of $\tau$ originating from the Higgs, in
particular in $\tau \rightarrow \pi \nu$, are expected to be boosted into the $\tau$ 
direction. Furthermore, the W and the top mass are
reconstructed, and the b-jet from the top quark decay is tagged.
More over a selection on missing $E_T$ is applied also in this case.
Figure~\ref{fig:hp5sigma} 
summarises the expected 5 $\sigma$  discovery reach for 30 $fb^{-1}$ for 
maximal stop mixing~\cite{mssm:hp5sigma}. 
A significant signal is expected for $\tan \beta >$ 15 between 180 GeV $\le m_A \le$.
The sensitivity for $m_A >$ 400 GeV  is for $\tan \beta >$ 25
Almost all the
parameter plane can be covered with the exception of a small area at low mA and low 
$\tan \beta$. 
\section{Triggering $\tau$ final states}
In order to permit the study of channels with $\tau$ final states a dedicated trigger system has been
studied. 
This trigger system is designed to be used in the selection of isolated $\tau$ leptons such as 
those expected in the MSSM Higgs decays $A/H \rightarrow \tau \tau$ and $H^{\pm} \rightarrow\tau \nu$. 
These include events with a lepton plus a $\tau_{jet}$,
two $\tau_{jet}$ and single $\tau_{jet}$ final states. 
The identification of $\tau_{jet}$ at trigger level is based on the isolation criteria made with the 
reconstructed tracks. 
Calorimeter triggers will provide a region to search for an isolated groups of tracks, 
well matched to the jet axis given by the calorimeter. 
Two different approaches can been used: one reconstructs tracks using only the inner
part of the tracking detector~\cite{trigger:sasha}, the other  uses tracking 
made with also the outer part of the tracking detector~\cite{trigger:tau}.
For the $H/A \rightarrow \tau \tau$ with hadronic $\tau$ decays both of them can be used, the first
is fastest while the second has bigger efficiency on signal events.
As already mentioned isolation criteria can reduce QCD contamination by a factor of 1000 when applied
to both jets.
For the $H^{\pm} \rightarrow \tau \nu$ channel and the other channels 
with one $\tau$ decayng leptonically
only the second selection can be used because it is the only one that can achieve the desired background
rejection.
These algorithms have satisfactory efficiency on signal events within the timing specification required
by a trigger apparatus.
\section{Conclusions}
In the MSSM, the almost the entire Higgs sector parameter space can be covered with 30 $fb^{-1}$
by the LHC experiments.
In many areas,
several Higgs bosons and decay modes will be available. 
The parameter choice for the MSSM physics studies is unavoidably restricted. 
The aim is to study a representative
set of parameters, and the detector performance and analysis lessons learned from the 
MSSM studies will serve to
explore any non-MSSM scenario.
In conclusion, as elusive as it is now, the Higgs sector will be well known, or well 
constrained, in seven years from now.
\section{Acknowledgements}
This report relies on the enormous amount of work done by my CMS colleagues. 
I thank especially Daniel Denegri, Sasha Nikitenko, Ritva Kinnunen, K. Lassilla-Perini and Giuseppe
Bagliesi
for useful advice and for providing the illustration.

\end{document}